\documentclass[a4paper]{article}
\usepackage{epsfig}

\usepackage{amsfonts}
\usepackage{bbm}

\begin{document}

%\doublespacing
%\flushbottom
%\raggedbottom
%\pagenumbering{roman}

\title{Relationalism vs. Bayesianism}\author{Thomas Marlow\thanks{email: pmxtm@nottingham.ac.uk}\\ \emph{School of Mathematical Sciences, University of Nottingham,}\\
\emph{UK, NG7 2RD}}

\maketitle

\begin{abstract}

We compare and contrast the basic principles of two philosophies:  Bayesianism and relationalism.  These two philosophies are both based upon criteria of rationality.  The analogy invoked in such a comparison seems rather apt when discussing tentative proofs of quantum nonlocality.  We argue that Bayesianism is almost to quantum theory, what general covariance is to general relativity.  This is because the Bayesian interpretation of quantum theory can be given a relational flavour.

\end{abstract}



\section*{Introduction}

In the following paper we suggest a curious analogy between two philosophies: relationalism and Bayesianism.  Basically there seems to be na\"{\i}ve connection between these two philosophies because the main principles have a similar wording.  It is obviously rather facetious just to equate the two philosophies for such a reason but, in the following, we wish to discuss where the analogy succeeds and where it fails.

According to Leibniz, a relational philosophy has to obey two principles of rationality (see \cite{Smolin05} for and accessible introduction).  Firstly we have the principle of sufficient reason which means that there must be a rational reason for invoking any feature of the theory.  This principle is intimately connected with the second; namely, the principle of identifying the indiscernible which states that if you have no rational reason to distinguish features of the theory you should identify them.  For example, in a relational theory of spacetime there is no discernable way to differentiate spacetimes which are shifted to the left by a certain amount, thus we identify all spacetimes that are related by such transformations.

Bayesian probability \cite{JaynesBOOK} has recently been invoked in a lot of physics circles.  Arguably, the basic principle of Bayesian probability theory is the principle of insufficient reason which states that if you cannot rationally prefer one of two propositions above the other you should assign them equal probabilities.  If you \emph{can} rationally promote one proposition of the two then you should assign them probabilities according to certain other rules that ensure they obey criteria of rationality \emph{i.e.} Cox's axioms of probability \cite{CoxBOOK}.  It is perhaps quite facetious to attempt to identify these two philosophies, but there is certainly a curious analogy between the two.  One might argue that the reason we assign the same probability to the faces of a fair die is because by believing it is fair, and given a sufficiently `good' randomising tossing procedure, we have no way to discern which face will come up top.

So the major principles of Bayesianism and relationalism are constraints due to rationality.  The question we wish to ask is whether there is anything to this curious analogy, or whether it is just a meaningless correlation in words.  The major reason we want to ask this question is because in searching for a quantum gravity theory we would like to keep the relationalism of general relativity and the probabilistic aspects of quantum theory (which, arguably, can be interpreted in a Bayesian manner \cite{Marlow06,Marlow05b,Fuchs02,Mana04,Srednick05}).

There are a few recent papers that hint about a relationship between these two philosophies \cite{Catich05,Poulin05,BH05}.  In \cite{Catich05} Caticha attempts to justify the thesis that a lot of the geometrical features of general relativity theories can be derived simply through a Bayesian theory of statistical inference.  In \cite{Poulin05} Poulin shows that the Bayesian interpretation of states is compatible with a variety of relationalism which then can be used to derive objects akin to spin-foams simply from a relational invocation of quantum theory.  In \cite{BH05} Bosse and Hartle discuss an attempt towards a covariant notion of a sum-over-histories interpretation which may also be interpreted in a rather Bayesian manner.

So, let us briefly introduce Bayesianism in a bit more detail to see if we can tease out any features of our tentative analogy between these two philosophies.

Bayesianism is a way of invoking, and deriving, a notion of probability simply from criteria of rationality.  In the standard analysis one can presume that the algebra of propositions has a Boolean structure and then invoke certain axioms that a rational notion of probabilistic inference should obey.  Using these axioms one can \emph{derive} the standard notions of probability and, furthermore, one can also derive relative frequencies with added assumptions also.  When, upon a given hypothesis, one does have a rational reason for interpreting one proposition as more probable than another then one should assign probabilities that reflect such a fact in a rational manner.  When we say `rational' we simply mean that one should assign them in such a way that we do not disobey certain criteria that our assignment should obey to be consistent.  Firstly we should presume that the probability we assign to the negation of a proposition upon a given hypothesis should only be a function of the probability of that proposition upon the same hypothesis.  This is Cox's first axiom of probability and can be written schematically as:

\begin{equation}
p(\neg \alpha \vert I) := G[p(\alpha \vert I)]
\label{COX2}
\end{equation}

\noindent where $G$ is an arbitrary function to be determined that is sufficiently well-behaved for our purposes and $I$ is the symbol we use for a hypothesis.  One can interpret this is a relational manner.  One does not wish to invoke any functional relations between probabilities that we cannot justify.  Our hypothesis $I$ should incorporate sufficient information for us to assign probabilities.  Of course this hypothesis could be false but we do not assign probabilities upon the presumption that it is false.  We assign probabilities upon the presumption that it is true.  If (\ref{COX2}) were not obeyed by a probability assignment then we would effectively be presuming that $I$ is not a good hypothesis, but we must assign probabilities by presuming it is a good hypothesis---otherwise what is the point?  If it is a bad hypothesis then we will infer probabilities that are inconsistent with what we observe and we may learn something from such a situation and assign new probabilities upon a new better `good' hypothesis \cite{JaynesBOOK}.

Similarly Cox's second axiom is a principle of rationality for similar reasons:

\begin{equation}
p(\alpha \cap \beta \vert I) := F[p(\alpha \vert \beta I), p(\beta \vert I)].
\label{COX1}
\end{equation}

We cannot justify any further functional relations between probabilities as these two axioms are sufficient to prove how probabilities should behave when using the $\cup$ operation.  From these two axioms and the presumption that probability assignments should be real (for arguments against this presumption see \cite{Marlow06}) we can easily derive the basic rules of probability theory:  probabilities obey Bayes' rule, and disjoint propositions have additive probabilities \cite{CoxBOOK}.  These, however, are rules about manipulating probabilities; they often do not constrain exactly what value we should assign as a probability.  For that we often need the principle of insufficient reason:  when, through our hypothesis, we have no rational reason to discern certain propositions we should assign them equal probabilities.

Thus one difference between the Bayesian principle of insufficient reason and the relational principle of identifying the indiscernible is the indirect nature of the Bayesian principle.  The principle of insufficient reason is about identifying the probabilities of propositions rather than the propositions themselves.  The principle of identifying the indiscernible is about treating propositions as equal rather than their probabilities.  Both principles are about not making any inference, or not justifying any result, based upon any spurious distinction between equals.  So we choose to have a background independent theory of gravity simply because we do not want our results to depend upon the vagaries of a background that we happen to choose.  Thus background independence is only a physical postulate in the sense that our physics should be rational.

One interesting distinction that must be made in relational theories of gravity is the distinction between \emph{active} and \emph{passive} diffeomorphisms \cite{Norton93}.  One chooses to invoke a generally covariant theory because such a theory `adds the least' to the catalogue of spacetime coincidences---it is explicitly a criteria of rationality.  But general covariance in the passive sense is just a mathematical property of certain theories, whereas it is the active general covariance that is meaningful physically through the above criteria of rationality.

So, we identify probabilities if we do not want our inferences to spuriously depend upon the fact we distinguish probabilistically propositions that we should not rationally.  We identify backgrounds if we cannot rationally justify one over the other so that any result we derive does not depend upon the vagaries of one particular background.  Of course, in the future, if we were to ever be able to distinguish one background over another we would change this assumption.  Similarly if we substitute one hypothesis for another our use of the principle of insufficient reason may change.  So both are principles of rationality ensuring that any physical/probable inferences we make are rationally justified---such inference is only as good as the original hypothesis.

Should acknowledging such a similarity between these two philosophies help in searching for a theory of quantum gravity?  Einstein's relational theory of gravity only involves inferences that are independent of any background spacetime we invoke.  The Bayesian theory of probability involves \emph{probabilistic} inferences. So perhaps the task of a quantum gravity theory is to find a theory in which we have probabilistic inference and background independent inference in the same theory.  Easily said, but not so easily done.

We are heartened by Hardy's recent `causaloid' work \cite{Hardy05}, which suggests that such a combination of such philosophies may indeed be possible (although it is still a long way off).  Hardy's work is quasi-operational (he invokes operationalism for practical reasons rather than necessarily accepting it as a philosophy) and invokes an operational notion of space-time that involves regions and actions upon these regions.  Such regions and actions are invoked using `cards' of data which are implicitly Boolean in nature (\emph{cf.} Bohr's philosophy and Einstein's notion of spacetime coincidence) and are combined in standard Boolean ways which suggests that Cox's proofs can be passed across and that one can invoke Bayes' rule (which Hardy does).  Hardy's formalism already incorporates quantum theory, should be able to incorporate general relativity, and tentatively should be able to incorporate a quantum gravity theory also.

How might we go about searching for such a relational quantum theory (also see \cite{Rovel96,Jaros05} for such relational ideas)?  In acknowledging the distinction between the two philosophies it seems we should first invoke a background independent set of propositions, and then use Bayesian reasoning to assign probabilities to such propositions.  Bayesianism should only be invoked once we have already identified the indiscernible---we must identify indiscernible propositions and then identify probabilistically those discernable propositions we cannot differentiate inferentially using a Bayesian hypothesis (and quantitatively differentiate---using a probability---those propositions we feel we can upon our hypothesis).  Obviously it is pure speculation to suggest that Bayesian reasoning is totally sufficient in itself to assign probabilities in this manner, but it can surely take us some of the way.

Bayesian probabilities may be useful in discussions of the histories propositional algebra \cite{Marlow06,Marlow05b} which is at least a small step towards such a generalisation.  In the case of the HPO history propositional algebra \cite{Isham94}, Bayesian probabilities seem to appear quite naturally. It is clear that von Neumann measurements are compatible with Bayesian probabilities but there is no \emph{a priori} reason why the history algebra should be compatible with Bayesian probabilities.  That there exist Bayesian assignments (linearly positive probabilities \cite{GP95} and complex probabilities \cite{Marlow06}) consistent with the HPO algebra is a surprising and suggestive result, especially since histories are often invoked as a small step towards a quantum gravity theory.

\section*{Why Use Bayesian Probability?}

If we are to invoke the above reasoning in any search for a quantum gravity theory it is prudent to explicitly place quantum theory in a Bayesian framework.  There are very good reasons for doing so even regardless of the above argument, and in this section we wish to outline a couple of said reasons.  It is clear that the measurement problem is nullified in a Bayesian framework because assignments are necessarily things that we assign.  Another problem that seems to be nullified is quantum nonlocality.

It is often stated that EPR and Bell-like theorems are avoided (not disproved but just avoided) by using Bayesian probability theory (this is often the major reason people invoke Bayesian reasoning in quantum theory \cite{Fuchs02}), but rarely is it explicitly stated how the invocation of Bayesian probabilities overcomes arguments for causal nonlocalities.  We shall attempt to argue why this is the case and why arguments that claim the opposite are lacking.

Bell proved that no hidden variable model (that obeys intuitive assumptions) can be used to emulate the correlations we find in certain correlated quantum systems \cite{BellBOOK}.  If we postulate variables that pre-exist measurement that aren't effected by measurement or actions happening at spacelike separation then we \emph{can't} use such theories to emulate quantum theory.  This is why many refer to an implicit `nonlocality' in quantum theory.  To put it another way, he proved that quantum theory cannot be completed by using such hidden variables.  One can also validly argue that Bell proved that quantum theory itself cannot obey those same intuitive assumptions regardless of hidden variables, so it either embodies causal nonlocalities or is not a \emph{hidden variable} completion of itself.  This is a result that goes against all our na\"{\i}ve assumptions about how the world works---a paradigm shifting body of work.

Of course, it is commonly accepted that Bell's analysis doesn't prove that quantum theory necessarily embodies \emph{causal} nonlocalities because it is not necessarily claimed that quantum theory is its own hidden variable completion.  This is easily seen by analysing the assumptions that Bell---happily acknowledging that one might make other assumptions---made.  The major assumption that Bell invoked is that the joint probabilities of outcomes at spacelike separation factorise in a certain way.  Namely, if we have two measurement parameters labelled $a$ and $b$, with outcomes $A$ and $B$ respectively, that are chosen at spacelike separation, then Bell assumes the following:

\begin{equation}
p(AB \vert ab\lambda) = p(A \vert a\lambda)p(B \vert b\lambda)
\label{locality}
\end{equation}

\noindent  where $\lambda$ are the hidden variables that we mentioned.  Bell called this factorisation assumption `local causality'.

Jarrett \cite{Jarrett84} showed that this factorisation assumption can be split up into two logically distinct assumptions which, when take together, imply Eq.\,(\ref{locality}). These two assumptions are, as named by Shimony \cite{Shimon86}, parameter independence and outcome independence.  Shimony showed that quantum theory itself obeys parameter independence, namely:

\begin{equation}
p(A \vert abI) = p(A \vert aI).
\label{parameter}
\end{equation}

\noindent
Here, $I$ represents all other information we have about the set-up including any hidden variables we invoke.  This means that, in quantum theory, the probability that one predicts for an event $A$ in one spacetime region does not change when the parameter $b$ chosen in measuring an entangled subsystem elsewhere is known. Knowledge of $b$ doesn't help in predicting anything about $A$.  But quantum theory does not obey outcome independence, namely:

\begin{equation}
p(A \vert BabI) \neq p(A \vert abI).
\label{outcome}
\end{equation}

\noindent
Knowledge of $B$ can and does affect the predictions one makes about $A$.  Thus, Bell's theorem proves that no parameter independent hidden variable theory that is also outcome independent can be used to emulate the correlations we get in quantum theory.  Of course, it is natural to assume outcome independence for hidden variable theories with no common causes.  Thus Bell made this assumption and proved his wonderfully iconic theorem that has now been experimentally verified over and over again.

Using notions of Bayesian theory \cite{JaynesBOOK} we can show concisely and explicitly why the presumption that quantum theory is its own hidden variable completion is not valid (thus blocking any proof that quantum theory, rather than its tentative completions, is a nonlocal theory).  Below we will look at a simple experiment---not intended as an emulation of quantum theory---which is outcome dependent and which cannot be described as a complete hidden variable model due to a \emph{physical} constraint upon what we are allowed to say about the variables involved (in a manner akin to the toy model in \cite{Spekkens04}).  Since such a constraint is justified physically we call such a model `complete'---it is the best theory we can rationally make of the physics.  We cannot complete the model further, in any manner akin to Bell's hidden variables, not because such variables would be non-local, but because it is inconsistent, or at least unjustified, physically to do so in the first place.

So, take an urn which is half-full of blue balls and half-full of red balls.  This urn is split into two sub-urns which each contain half the total number of balls in the original urn.  The selection process which chooses which balls to put in which sub-urn is independent of the colour of the balls.  Thus if we take one ball out of a sub-urn and note that it is a red ball it is clear that, due to the finite number of balls, the probability we now predict (given that further information) of another observer taking a red-ball from the other sub-urn---upon his first dip into the other sub-urn---is diminished. This classical result is true regardless of the possible spacelike separation between the sub-urns. When the other observer and ourselves are spacelike separated then the predictions that the other observer makes will be unchanged by our action because he cannot receive any information from us due to special relativistic considerations, but, still, such an experiment is outcome dependent.  Everyone must surely agree that all mechanisms involved here are locally-causal and yet the two sub-urns form an outcome dependent experiment.  It is clear that we don't propose a mechanical effect between the two sub-urns.  Nor would anyone claim that the reality of the balls we use is in doubt; we only doubt that we can know the configuration of the sub-urns without changing the nature of the experiment (making the sub-urns transparent say)---presuming that we do know the contents of the sub-urns but are ignoring them is not a valid assumption because, by the very nature of the experiment, we \emph{do not} know the contents of the sub-urns---some other demon observer who is actually in the sub-urns might know the correct configuration, but we certainly do not (we quantify our ignorance not their's).  Rationally, we should differentiate the invocation of demon observers from the invocation of a single no-nonsense observer.  If we were to introduce two more properties, say a magnetic or non-magnetic property of the balls then, observers of the two sub-urns could freely decide whether to measure colour or magnetic properties of balls they pick out, but still we could get some form of outcome dependence.

Of course, in Bell's argument, the whole point of presuming outcome independence is because \emph{in presuming} hidden variables (or equivalently demon observers) these hidden variables should complete our notion of state such that the probabilities do factorise in the manner that Bell invokes.  Let us state clearly that we do not have any issue at all with Bell's proof---we take it as utterly incontrovertible.  Bell's proof shows that such completions \emph{do not give a philosophically appealing theory}---they are nonlocal.  One might attempt to argue that because quantum probabilities do not obey outcome independence \emph{even regardless of appended hidden variables}, then it must embody such causal nonlocalities.  This argument, however, does not follow through, because we do not presume that the variables involved ($a,b,A,B$) are hidden variables that complete quantum theory.  As soon as one interprets probabilities using an ignorance interpretation then it is clear that outcome dependence is a possibility in exactly the manner akin to our two-sub-urns experiment.  As such, we do not have any qualms as of yet about nonlocality in quantum theory.  We shall discuss Bell-like theorems which do not use probabilistic assumptions later; but first let us carefully discuss why using an ignorance interpretation of probabilities in quantum theory is useful.

Consider the analogy with general covariance.  We invoke general covariance because we do not want our inferences to depend upon the vagaries of a background spacetime that we happen to choose.  Similarly we invoke Bayesian probability because we do not want \emph{our} inferences (rather than a demon observer's) to depend upon the vagaries of a particular average over hidden configurations that we might presume to be the case.  Therefore we would rather call quantum theory `complete' than `incomplete'.  It is as complete as any rational notion can be.  Just as a background would complete a background independent theory, so would a hidden variable theory complete quantum mechanics.  However, a background independent theory is more physically appealing than any tentative not-rationally-justified completion.  We argue for an analogous account for hidden variable completions of quantum theory.  When we wish to differentiate Bell's use of the word complete from ours we shall write our notion of `complete' within scare quotes.

Of course, outcome dependence or independence in quantum theory is defined with respect to the whole state of the system and apparatus.  But, for example, the state that we assign to the system and apparatus in our Bayesian two-sub-urns example cannot include a specification of the numbers of each colour in each half-urn because we do not know such a specification.  However, as soon as we begin to accept that the state we assign to the system plus apparatus does not, even implicitly, contain such a specification hidden from us (why should it because we don't know the specification and don't, and arguably shouldn't, build it into the theory because it is a rational theory of our ignorance) then we can still call the state `complete' in a limited sense.  The specification of the contents of the sub-urns is unknown.  A `complete' state of the system and apparatus should acknowledge this fact.  It happens to be the case that in our Bayesian two-sub-urns example that our prior-knowledge of the set-up allows us to design a hidden variable theory because, relative to the probabilistic state we assign the system plus apparatus, there is only one possible list of counterfactually possible configurations.  There is no \emph{a priori} reason this should be the case;  equivalence between ignorance theories and hidden variable theories might happen to be so in certain theories but we have no reason to assume it is the case.  In quantum theory it is explicitly not the case because our prior-knowledge of the system and apparatus is not sufficient for us to infer a unique hidden variable theory.

So, one likes to assume that some other demon observer knowing the real `hidden' configuration of the sub-urns does not affect the predictions we make regarding the system and apparatus based upon the fact that we do not know the hidden configuration of the sub-urns.  This we do not argue against.  Of course, knowing the hidden configuration (or outcomes in certain arms of the experiment) does mean that the demon's probabilistic predictions regarding the system and apparatus are manifestly different from ours.

Say we have a demon observer who knows a particular configuration is correct.  Equally, probabilistically speaking, the configuration could have been different (there are many hidden configurations that are possible for a given state assigned by an observer who doesn't know the hidden configuration---who is ignorant of it).  We can then tentatively list \emph{the} set of all counterfactually distinct possible configurations, each of which is known by a counterfactually distinct demon observer.  Normally ignorance is associated with an ignorance as to what counterfactually distinct possibility is actually the case.  But, why should we presume that such a set of counterfactually distinct demon observers who each know one of the possible hidden configurations is operationally equivalent to one observer who does not know the configuration?  What justifies such an assumption?  Knowing the hidden configuration changes one's predictions.  Demon observers who know the configuration will predict different probabilities for future (or past) events in comparison to an observer who does not know the configuration (it is `hidden' from her).  So why presume that a list of demon observers \emph{who each predict different things to the ignorant observer} is operationally equivalent to the ignorant observer?  This is an assumption that has not been rationally or operationally justified.  It is implicitly assumed in the normal ensemble interpretation of probabilities.

Knowledge changes one's predictions.  It is only in ignorance that one maintains all one's predictions.  There is no rational reason to equate ignorance with the set of all counterfactually possible cases of non-ignorance.  Thus there are distinct ways that we can interpret ignorance.  To distinguish such interpretations we shall call the unknown configuration (naturally, being a realist, we presume that such a configuration is the case, we just cannot justify representing it in a theory of our ignorance) presumed but not modelled by an ignorant observer simply the `unknown configuration', and we shall call a configuration that is known by some other demon observer a `hidden configuration' (it is hidden relative to an ignorant observer).  Bell proved that we cannot presume a hidden configuration in quantum theory without also having to acknowledge some form of nonlocality.  His proof, of course, does not apply to an unknown configuration---the proof does not follow through except for the hidden configurations which Bell considered.  So one might argue against the pedagogical use of demon observers rather than argue for nonlocality in quantum theory.  In the least, we shall learn something in the process of attempting such a task.

Is there an implicit problem in invoking a list of counterfactually distinct demon observers?  Demon observers are simply not operationally justified, just as a particular background spacetime is not operationally justified.  Perhaps we can presume that we could have known the `real' configuration, but we certainly cannot presume that we \emph{do} know the `real' configuration because we are presuming ignorance.  Should we instead use the standard Bayesian notion of ignorance, based on the ignorance of a \emph{single} observer?  In the standard Bayesian notion, an observer's ignorance affects the probabilities \emph{he assigns}.  His assignments are explicitly not effected by, and rationally should not depend upon, the probabilities that other uncommunicative observers assign.  When we use such an ignorance interpretation we are allowed outcome dependence in such a manner that it does not necessarily conflict with special relativity just as we are allowed outcome dependence in an ignorance interpretation of the two-sub-urns experiment.  The thesis that using Bayesian reasoning in Bell's analysis nullifies any proof that quantum mechanics (rather than quantum mechanics appended with hidden configurations) embodies causal nonlocalities is commonly hinted at in the literature (it is often invoked as a reason for choosing to interpret quantum probabilities in such a manner), but rarely is is it expressed explicitly.  Bell's theorem tells us that either quantum theory is nonlocal or it (rather than its tentative completions) is a theory which we are not justified in calling a hidden variable theory.  Clearly the latter proposal is far less drastic than the former because we can interpret quantum theory as a theory of quantified ignorance.

Of course, we have taken liberties with our use of the term `ignorant'.  An observer who assigns a state to a system plus apparatus in quantum theory is really not that ignorant at all.  He can predict probabilities for any von Neumann measurement (or POVM) he would like to make.  His inferences, which he quantifies using probabilities, may, of course, be wrong.  When using the term ignorant above we obviously only mean that he is comparatively ignorant in comparison to a postulated demon observer who does know the hidden configuration.  So, we can postulate demon observers and equate ignorance with a lack of knowledge as to which demon is correct.  Then, according to Bell, we get nonlocality.  Clearly a meta-demon who is ignorant of which demon minion is correct knows a lot more than we do---he knows which list of counterfactually possible configurations to use and what prior-probabilities to assign them.  In quantum theory, we simply do not.  It is clear that in quantum theory we can imagine many different lists of counterfactually distinct hidden configurations---there are many hidden variable theories.  Why rationally choose one over the other?

The nice property of the Bayesian interpretation of ignorance is that it is, perhaps ironically, not particularly cyclical in its reasoning.  We model an ignorant observer's probability assignments using the standard criteria of rationality: the principle of insufficient reason.  If, however, we model ignorance as the ignorance of a meta-demon observer who is unclear as to which of his demon minions is correct then we need lots of prior notions of probability.  The Bayesian notion is easy to understand and does not add any spurious assumptions to the theory---it is a criteria of rationality just as general covariance is; it is also as physical an assumption as general covariance is.  Such a model would only be `complete' in the sense that we do not invoke anything we are not rationally compelled to.  No further completion is rationally justified.  Each and every background dependent theory being irrationally justified does not mean that a background independent theory is irrational---quite the opposite, a background independent theory is clearly not the sum total of all background dependent theories.  Similarly with hidden variable theories and a Bayesian interpretation of quantum mechanics.  It is comonly accepted that some hidden variable theories require a special foliation or frame of reference, so the analogy may not be as off-the-mark as one might initially consider it.  So---we intend this only half-seriously---perhaps Bayesian quantum theory is already a partially background independent theory (or at least should accommodate one to some extent).

Thus outcome dependence does not \emph{necessarily} prove anything about nonlocal causality. The colour of each ball, or the contents of each sub-urn, is not necessarily a hidden (what we might rather call a known-but-ignored) variable \emph{per se}, but rather, we argue, it is an unknown variable.  We quantify how unknown it is by a probability.  (We do not use probability to quantify how known something is to some postulated other demon observer, we use it to quantify how known something is to \emph{us}.)  This does not nullify Bell's proof, but rather it nullifies any suggestion that quantum theory is itself a hidden variable theory upon which we can apply Bell's proof.  Ignorance, in the Bayesian interpretation, is not about meta-demons, it is about a standard no-nonsense observer.  Even though our two-sub-urns experiment \emph{can} be given a hidden variable formulation, interpreting it in a Bayesian fashion removes the justification for doing so in the first place.  If one doesn't like the cyclical nature of invoking prior-probabilities and demons to define probabilities in quantum theory then don't use them; use Bayesian principles.

It is perhaps rather ironic that we have had to invoke an interpretation of probability based around `ignorance' in order to discuss the tentative `completeness' of quantum theory.  One might query: ``Surely, by definition, a theory of quantified ignorance is necessarily incomplete?''.  Yes, but if something is \emph{necessarily} incomplete then why attempt to complete it?  And, why care that such a completion should behave nicely if one does?  One can never complete our two-sub-urns experiment because it tells us everything it can and any natural completion requires a change in the very predictions that we make (making the urns transparent say).  Therefore it is the best physical description we can ever get while making the same predictions, hence it is natural to call it a `complete' description of an unknown realistic configuration.  If a theory \emph{cannot} be rationally completed while retaining its predictions then it is either `complete' or makes the wrong predictions.  Therefore we do not reject the idea that a new theory might come about that will supersede quantum theory---we only reject, like Bell, that any natural completion of quantum theory cannot obey Bell locality.  This is not \emph{necessarily} in conflict with quantum theory being `complete' (in the same sense that our two-sub-urn predictions are `complete' in a Bayesian interpretation) nor with it being entirely consistent with special relativity and the free choice of observers---this is clear by analogy with our Bayesian two-sub-urns example; a simple `complete' description of an unknown configuration of realistic balls.  A complete description of a known configuration of realistic balls is a whole different kettle of fish and a complete description of an unknown configuration is simply a misnomer.  In order to tentatively deny any conflicts between special relativity and quantum mechanics one would probably have to use something akin to Bayesian probability.  One can refuse to take this path if it is not aesthetically pleasing to you, but one should not, yet, deny its existence.  Just as Eq.(\ref{locality}) ``may not embody \emph{your} notion of local causality'' \cite{BellBOOKsub}, neither might our notion of `completeness' embody your notion.  At present, there is sufficient ambiguity implicit in foundational notions of probability---and in all the other dubious words we use---to justify either deep truth: `nonlocality' or `completeness'.

One might then, na\"{\i}vely, ask whether such outcome dependent experiments can emulate the correlations we get in quantum theory.  But, of course, if we allow outcome dependence then we cannot even derive Bell's inequalities.  So the more pertinent question is, can such outcome dependent experiments emulate the probabilistic predictions of quantum theory?  Quantum theory is itself an outcome dependent theory so the possibility remains that---since quantum theory is itself a theory we use to predict the probabilities of unknown variables---quantum theory is its own `completion'.  We don't claim this view is new; it was, for example, hinted at by Jaynes in \cite{Jaynes89}---although in a rather contentious manner which is easily criticised \cite{Potvin04}.  Such a proposal is obviously analogous to stating that quantum theory is just a novel probability theory \cite{Hardy01,Youssef94}.

Obviously, we do \emph{not} claim that we can emulate quantum theory using such urns and balls; rather we argue only that accepting outcome dependence can be completely compatible with special relativistic causality and the free choice of observers is not an escape route that has been exhaustively proved wanting.

Such a proposal is akin to the standard way that most try to get around Bell's theorem---one denies the hidden variable hypothesis.  But, as we argue here, when denying the hidden variable hypothesis one need not deny all forms of realism.  When interpreted in a Bayesian fashion, certain simple urn experiments use a form of realism distinct from the realism used when invoking hidden variables---such experiments are still naturally considered eminently realistic.  Such realistic experiments also require outcome dependence in a manner that is not necessarily in any conflict with special relativity.  In our two-sub-urns example one might try to say that outcome dependence is, in this case, not incompatible with local causality \emph{because} we can design a local hidden variable theory, but such a claim is rather dubious.  In quantum theory outcome dependence might be compatible with local causality in a manner distinct from the manner in which our two-sub-urns example is---quantum theory simply isn't classical probability theory.  So perhaps quantum theory is a realistic outcome dependent theory where the specification of hidden variables in the theory is a \emph{physically} dubious assumption---and not just philosophically unappealing---just as background dependence is physically dubious because it is not rationally justified.  We call it `physically dubious' because it takes the form of a necessary constraint upon how we are allowed to consistently represent certain variables in physical theories.

So, such an analysis is not a rejection of realism \emph{per se}, it is rather just an acceptance of a certain realism above and beyond the na\"{\i}ve kind used when discussing hidden variables.  One is not to interpret the term `na\"{\i}ve' in any derogatory fashion.  This is, after all, exactly what Bell profoundly proved; that we cannot use such notions when discussing quantum theory, hence their na\"{\i}vety---the domain of his discussion is clearly stated.  So, lets not search for a completion of the theory (Bell tells us it is nonlocal and by the above argument such a task is not rationally justified) but rather lets search for a locally causal `complete' theory---`complete' in the sense that the state represents everything we can rationally justify that we know about the system and apparatus.  Bell's work is the first step one must take in such a search; one must reject a wide class of intuitive theories.  (Or, in opposition to this tentative view, one must search for a proof that quantum theory contains causal nonlocalities.)  The beauty of Bell's work is that such a first step could be taken by experimental verification.  Unfortunately we do not have the genius to propose a way to experimentally verify the second conceptual shift we have just invoked---using Bayesian probability theory.  Nethertheless, such a conceptual shift is perhaps warranted by the aesthetics of such an approach.

Having recognised the domain of Bell's work, many authors have attempted to extend and generalise the discussion without having to use the probabilistic assumptions that Bell used.  Some good examples are the exciting programmes started by Hardy \cite{Hardy92} and Stapp \cite{Stapp03}.  Stapp's programme, for example, is in opposition to the tentative view given above; it is an attempt to prove that causal nonlocalities are necessarily required by quantum theory.  Stapp's work, however, has recently been criticised by Shimony \cite{Shimon04}.  We argued above that, because of outcome dependence, probabilities are not well-defined independently of events in space-like separated regions---they are manifestly defined with respect to such regions.  (Note the distinction between a probability that is well-defined due to an ignorance of spacelike separated events and a probability that is independent of knowledge of events at spacelike separated events; the latter is ill-defined.) Similarly, Shimony argues that the counterfactual statements that Stapp invokes are also not well-defined independently of space-like separated regions---counterfactual statements are manifestly defined relative to such regions.  Such interdependencies between the very definitions of certain probabilities or between the very definitions of certain counterfactual statements may be interpreted in an entirely logical manner---thus we need not \emph{necessarily} invoke nonlocal causal relations to explain them.  Thus it is not yet clear that such programmes prove that causal nonlocalities are necessarily a component of quantum theory.  Of course, all this relates to the perennial debate started by the conflict between the EPR paper \cite{EPR} and Bohr's response \cite{Bohr35}.  All we argue here is the debate is not concluded.

Depending on your personal views on the foundations of probability one may or may not take to the above treatise for the use of Bayesian probability in physics.  Recently a spate of work has been written with such a Bayesian commitment.  Essential highlights are \cite{JaynesBOOK,Fuchs02,Mana04,Srednick05,Catich03}.  If, after reading such work, one does not accept the Bayesian view then one might also wish to take note of the recent discussion of Bell's inequalities by Khrennikov \cite{Khrennik02} written specifically in terms of von Mises' relative frequency approach \cite{MisesBOOK}.

Bayesian probabilities are invoked as subjective degrees of belief---but this is not incompatible with realism (also see \cite{Youssef01}).  When all observers agree upon probability assignments then the probabilities are intersubjectively defined---which is, in turn, as close to objectivity as we ever get.  In order to find such intersubjective probabilities we would have to discuss a subset of all the probabilities invoked by applying some kind of constraint---perhaps something akin to exchangability of sequences of experimental results.  Depending upon the constraint required one might be able to invoke a locally causal hidden variable model of such intersubjective probabilities---the most obvious constraint is outcome independence itself, so one, perhaps, doesn't have to do much work!  However, quantum theory can describe more than just these intersubjectively defined probabilities.

\section*{Bayesian Histories}

So, we argue that there are good reasons of making quantum theory explicitly a Bayesian theory.  We have recently suggested two possible ways this could be done while using a history formulation of quantum theory.  Firstly we have argued \cite{Marlow05b} that the Linearly Positive formalism of Goldstein and Page \cite{GP95} is explicitly compatible with Cox's Bayesian axioms of probability and secondly we have argued that a more general complex probability assignment justified using Cox's axioms also makes quite a lot of sense \cite{Marlow06}.  If one could find a constructive uniqueness proof for these Bayesian probability assignments we would be one step closer to a quantum gravity theory.  We would then have a derivation of Bayesian probabilities from a propositional algebra with quantum properties.  The next step would be to invoke an analogous generally covariant `histories' or `spacetimes' algebra and an analogous construction of Bayesian assignments.  Perhaps a covariant propositional algebra should involve whole spacetimes or causal sets rather than histories \emph{per se}.  General covariance is only justified as a criteria of rationality in discussing cosmological situations \emph{i.e.} whole spacetimes.  General covariance has a trivial symmetry group \cite{Norton93}.  When not discussing cosmological situations boundary conditions ensure that non-trivial symmetries should be invoked.  Such a non-cosmological theory should still be as background independent as the boundary conditions allow (we should not base inferences on any background structures \emph{except} the boundary conditions themselves).  This is something that is explicitly acknowledged in Hardy's framework too \cite{Hardy05}.

One might reject the use of Bayesian probabilities in cosmological situations because there are no observers `outside' the universe that we can invoke, but this is simply a misnomer.  Bayesian probability is widely used in the astrophysics community for example.

\section*{Conclusion}

So, we argue, there are good reasons for invoking Bayesian probabilities in quantum theory.  Interpreting quantum theory in a Bayesian manner explicitly stops us from necessarily having to invoke nonlocality, and Bayesian assignments seem to appear quite naturally within the quantum histories formalism---of course the histories algebra is not generally covariant and nor have we yet proved that our suggested Bayesian assignments are unique; but, arguably, the primary reason many physicists have invoked history theories is that it seems to be one step closer to a covariant formulation \cite{BH05,Isham94,GH90,Hartle04}.  Some even suggest that quantum theory is \emph{just} a novel probability theory that may be derivable from simple plausible premises \cite{Youssef01,Hardy01}.  If this is the case then the most natural programme for quantum gravity is just to apply this novel probability theory to a generally covariant propositional algebra.  Again, this is far easier said than done.  First we need to justify a generally covariant propositional algebra (this would incorporate the kinematics and dynamics of the theory).  Hardy \cite{Hardy05} has recently suggested a tentative mathematical framework in which such a theory could be defined---although Hardy doesn't explicitly use Bayesian probabilities (he does however use Bayes' formula), it seems that a notion of probability based around criteria of rationality might be exactly what we need for a relational quantum gravity.

\section*{Acknowledgements}

I'd like to kindly thank Brad Weslake for a discussion which helped clarify exactly what I was trying to say in regards to Bell-like theorems, and EPSRC for funding this work.

\end{document}